\documentclass[aps,rmp,twocolumn]{revtex4}
\usepackage{graphicx}
\usepackage{amsmath}
\usepackage{url}
\makeatletter
\@ifxundefined\bibnotes@sw{\@booleantrue\bibnotes@sw}{}%
\bibpunct{}{}{,}{s}{}{\textsuperscript{,}}%
\def\@onlinecite#1{\begingroup\let\@cite\NAT@citenum\citealp{#1}\endgroup}%
\appdef\class@documenthook{%
 \@ifxundefined\place@bibnumber{%
  \let\place@bibnumber\place@bibnumber@sup
 }{}%
}%
\makeatother
\begin{document}

\title{100 Years of Brownian motion}
\author{Peter H{\"a}nggi}

\affiliation{Institut f{\"u}r Physik, Universit{\"a}t Augsburg,
             86135 Augsburg, Germany}
\author{Fabio Marchesoni}
\affiliation{Dipartimento di Fisica, Universit\`a di Camerino,
I-62032 Camerino, Italy}

\date{\today}

\maketitle

\section{Introduction}
In the year 1905 Albert Einstein published four papers that raised
him to a giant in the  history of science of all times. These works
encompass the photon hypothesis (for which he obtained the Nobel
prize in 1921), his first two papers on (special) relativity theory
and, of course, his first paper on Brownian motion, entitled
``\"Uber die von der molekularkinetischen Theorie der W\"arme
geforderte Bewegung von in ruhenden Fl\"ussigkeiten suspendierten
Teilchen'' (submitted on May 11, 1905) \cite{einstein1905}. Thanks
to Einstein intuition, the phenomenon observed by the Scottish
botanist Rober Brown in 1827 \cite{brown1828} -- a little more than
a naturalist's curiosity -- becomes the keystone of a fully
probabilistic formulation of statistical mechanics and a
well-established subject of physical investigation which we
celebrate in this Focus issue entitled -- for this reason -- : ``100
Years of Brownian Motion''.

Although written in a dated language, Einstein's first paper on
Brownian motion already contains the cornerstones of the modern
theory of stochastic processes. The author starts out by using
arguments of thermodynamics and the concept of osmotic pressure of
suspended particles to evaluate a particle diffusion constant by
balancing a diffusion current with a drift current (through Stokes'
law). In doing so, he obtains a relation between two transport
coefficients: the particle diffusion constant and the fluid
viscosity, or friction. This relation, known as the Einstein
relation \cite{pais}, was generalized later on in terms of the
famous fluctuation-dissipation theorem by Callen and Welton
\cite{callenwelton1951}, and with the linear response theory by Kubo
\cite{kubo1957}. A much clearer discussion of Einstein's arguments
can be found in his thesis work, accepted by the University of
Zurich in July 1905, which he submitted for publication on August
19, 1905 \cite{einstein1906}.

The second part of his 1905 paper contains a heuristic derivation
of the (overdamped) diffusion equation,
from which he deduces his famous prediction
that the root mean square displacement of suspended particles is
proportional to the square root of time. Moreover, the trajectories
of a Brownian particle can be regarded as
{\it memory-less} and {\it non-differentiable} \cite{einstein1908},
so that its motion is {\it not} ballistic
(a bold statement that troubled mathematicians for half a century!).
The latter also explained why earlier
attempts to measure the velocity of Brownian particles yielded puzzling results,
and consequently were doomed to fail.

A crucial consequence of Einstein's theory is that
from a measurement of the diffusion constant
 -- i.e. by measuring {\it distance} traveled rather than velocity --
it would be possible to extract an independent estimate of the
atomistic important, and much debated Avogadro-Loschmidt number $N$.
Notably, the earliest determination of this number dates back to
1865 (!) when Johann Josef Loschmidt tried first to measure the size
of molecules \cite{bader01}: his data for mercury were compatible
with a ``best'' value of 4.4 $\times 10^{23}$ molecules per mole.
Inspired by Einstein's work, an ingenious ``reality check'' on the
role of fluctuations was performed through a series of experiments
by J. Perrin and his students in 1908--1911 \cite{perrin};
Einstein's predictions could be beautifully verified by setting the
Avogadro-Loschmidt number in the range (6.4 $\div$ 6.9)$\times
10^{23}$/[mol]; by 1914 the first three digits of the actual figure
of $6.022 1415 \times 10^{23}$/[mol] with a standard uncertainty of
$0.0000010 \times 10^{23}$/[mol], were finally accepted
\cite{virgo33}.

The publication of Einstein's papers provided further strong
evidence for the atomistic hypothesis of matter. The immediate
validation of his theory finally vindicated the arguments of the
``discontinuists''; the remaining ``continuists", such as Wilhelm
Ostwald, and in particular Ernst Mach [the latter being famous for
his cynical remark to all ``discontinuists'': ``haben wir's denn
gesehen?''(die Atome/Molek\"ule), meaning ``have we actually seen
it?''(the atoms/molecules)] had thus no choice left but to concede.

We will not belabor any further the history of Brownian motion and
the pioneering developments of its theory by
Einstein's contemporaries like Marian von Smoluchowski
\cite{smoluchowski1906}
(who worked on the molecular kinetic approach to Brownian motion since 1900,
but did not publish until 1906), Paul Langevin
\cite{langevin1908}, and Norbert Wiener \cite{wiener}.
Beautiful accounts have been given in the
literature by several authors.
We mention here in particular the intriguing and most insightful
introductory chapter by  R.M. Mazo
\cite{mazo2002},  the short histories by M.D. Haw \cite{haw2002}
and J.G. Powles \cite{powles78},
or the notes presented by E. Nelson \cite{nelson1967}.

\section{The impact of Brownian motion theory up to present}

Without any doubt, the problem of Brownian motion has played a central
role in the development of both the foundations of thermodynamics and the
dynamical interpretation of statistical physics.
A theory of Brownian motion based on the molecular-kinetic theory
of heat, as that proposed by Einstein in 1905, does provide the link
between an elementary underlying microscopic dynamics and
macroscopic observable phenomena, such as the ubiquitous fluctuations of extended
systems in natural and social science.

The early theories of Brownian motion inspired many prominent
developments in various areas of physics, still subject of active research.
In the following we briefly mention some of those addressed
in the present Focus issue.

Among the first to dwell on the ramifications of the fluctuation-dissipation
relation were, as mentioned already, Callen and Welton \cite{callenwelton1951}:
These authors generalized the relations
by Einstein, and subsequently by Nyquist and
Johnson for the voltage fluctuations, to include quantum effects.
In their fundamental work, they established a generally
valid connection between the response function and the associated equilibrium
quantum fluctuations, i.e. the quantum fluctuation-dissipation theorem.

Another key development must be credited to Lars Onsager: Via his
{\it regression hypothesis}, he linked the relaxation of an
observable in the presence of weak external perturbations to the
decay of correlations between associated microscopic variables
\cite{onsager1931}. This all culminated in the relations commonly
known as the Green-Kubo relations
\cite{kubo1957,green1952,kubo1966}. This notion of ``Linear
Response'' which in turn is related to the fluctuation properties of
the corresponding variables (response-fluctuation theorems) can as
well be extended to arbitrary (dynamical and non-dynamical) systems
that operate far from equilibrium \cite{hanggithomas1982}: The
corresponding fluctuation-theorem relations (where the imaginary
part of response function  generally  is no longer related to the
mechanism of physical energy-dissipation \cite{hanggithomas1982})
provide most valuable information on the role of {\it
non-equilibrium} fluctuations.

These classic ``fluctuation theorems'', which describe the linear
response to external perturbations in arbitrary
statistical systems far away from thermal equilibrium,
should not be confused with the recent beautiful
{\it non-equilibrium work relations}, often also termed {\it fluctuation theorems}.
This latter branch of fluctuation research was initiated by Evans {\it et al.} \cite{FT}
and then formalized in the {\it chaotic hypothesis} of Galavotti and Cohen \cite{FT}.
Independently, Jarzynski \cite{jarzynski97}
proposed an interesting {\it equality}, being valid for both
closed and open {\it classical}
statistical systems: It relates -- a priori surprisingly --
the difference of two {\it equilibrium} free energies to the
expectation of a particularly designed, stylized {\it non-equilibrium}
work functional.

There is also an ongoing debate on the true origin of irregularity
that causes the stochastic, random character
of Brownian trajectories. In particular, is a chaotic microscopic
dynamics sufficient, or
is it more the role played by the extreme high dimensionality of
the phase space that, on reduction,
causes the jittery motion of the individual Brownian particles?
The present Focus issue contains an elucidative
contribution by Vulpiani and collaborators \cite{vulpiani05}, who
address right this and related issues.
Answering this basic question becomes even more difficult when we attempt
to include quantum mechanics.

The description of Brownian motion for general quantum systems still
presents  true challenges, see the discussion
herein by H\"anggi and Ingold
\cite{hanggiingold05} and Ankerhold {\it et al.} \cite{ankerhold05}.
For example, little is
known for the modelling from first principles of quantum fluctuations
in stationary non-equilibrium systems,
nor on the connection between the complexity obtained upon phase-space reduction
and the microscopic quantum chaos.

The theory of Brownian motion also had a substantial impact on the
theory of quantum mechanics itself.
The formulation of quantum mechanics
as a sum over paths \cite{feynman1948,kleinert2004} has its roots in the
diffusive nature of the trajectories
of a Brownian walker in continuous time: The Feynman-Kac propagator
is nothing but a Schr\"odinger equation in
imaginary time. In diffusion theory this idea had been utilized as
early as in 1953 by Onsager and Machlup
\cite{onsagermachlup1953} for Gauss-Markov processes with linear coefficients.
Its nontrivial extensions to the case with nonlinear drifts and nonlinear
diffusion coefficients \cite{piFP} and  to colored noise driven nonlinear dynamics \cite{piCOLOR},
have been mastered only 15-30 years ago .

The debate on Brownian motion also inspired mathematicians like
Cauchy, Khintchine, L{\'e}vy, Mandelbrot,
and many physicists and engineers to go beyond Einstein's formulation.
Non-differentiable Brownian trajectories
in modern language
are called ``fractal''
and statistically self similar on all scales.  These extensions carry
names such as fractal Brownian motion, L{\'e}vy noise,
L{\'e}vy flights, L{\'e}vy walks, continuous time random walks,
fractal diffusion, etc., \cite{fractalbm,feder88}.
This topic is presently of wide interest and is being used to
describe a variety of complex
physical phenomena exhibiting e.g.
the anomalous diffusive behaviors reviewed here by Sokolov and Klafter
\cite{sokolovklafter05}, or the
diffusion limited growth and aggregation mechanisms discussed by
Sander and Somfai \cite{sandersomfai05}.

The quest for a mathematical description of the Brownian trajectories
led to a new class of differential equations, namely the so-called
stochastic differential equations. Such equations can be regarded
as generalizations -- pioneered by Paul Langevin -- of Newtonian
mechanical equations that are
driven by independent, stochastic increments obeying either a Gaussian
(white Gaussian noise) or a Poisson statistics
(white Poisson noise). This yields a formulation of the Fokker-Planck equations
(master equations) in terms of a nonlinear
Langevin equations generally driven by multiplicative,
white Gaussian (Poisson) noise(s). As
the aforementioned independent increments correspond to no bounded
trajectory variations,
the integration of such differential equations
must be given a more general meaning:
This led to the stochastic integration calculus of either the Ito type,
the Stratonovich type, or generalizations thereof
\cite{hanggithomas1982,risken92,coffey04}.
In recent years, this method of modelling the statistical mechanics
of generally nonlinear systems driven
by random forces has been developed further to account for physically more realistic noise
sources possessing a finite or even infinite
noise-correlation time (colored noise), i.e. noise sources that are
non-Markovian \cite{hanggijung1995}. In this Focus issue Luczka \cite{luczka05}
provides a timely overview of this recent progress together with newest findings.

A powerful scheme to describe and characterize a statistical nonlinear dynamics
from microscopic first principles is given by the methodology of non-Markovian,
generalized Langevin equations or its associated
generalized master equations. This strategy is by now well
developed and understood only for \textit{thermal equilibrium systems}. The projector operator
approach \cite{GME,zwanzig}, which is used to eliminate the non-relevant
(phase space) degrees of freedom, yields a
clear-cut method to obtain the formal equations for either the rate of change
of the probability or the reduced density operator,
i.e. the generalized (quantum) master equation or the nonlinear generalized (quantum) Langevin
equation \cite{GLE}. This latter approach proved very useful to characterize the
complex relaxation dynamics in glasses and related systems \cite{glassphysics}.

There exists an abundance of processes in physics, chemistry
(chemical kinetics), biology and engineering, where the dynamics
involves activated barrier crossings and/or quantum
tunnelling-assisted processes through barriers. In all these
processes the time-scale for escape events is governed by
fluctuations that typically are of Brownian motion origin. The first
attempts to characterize escape dynamics date back to the early
thirties with contributions by Farkas, Wigner, Eyring, Kramers, to
name a few prominent ones. This topic has been extended in the late
seventies early eighties to account also for (non-Markovian) memory
effects, solvent effects, quantum tunnelling, non-equilibrium
fluctuations, correlated noises (i.e. colored noises)
\cite{hanggijung1995}, nonlinear bath degrees of freedom and
time-dependent forcing. The interested reader is directed to a
comprehensive review \cite{rmp90} and is further referred to the
up-to-date accounts given by Pollak and Talkner
\cite{pollaktalkner05} and H\"anggi and Ingold \cite{hanggiingold05}
in this issue.

The combined action of external driving and noise has given rise
to new phenomena, where the constructive role of Brownian motion
provides a rich scenario of far-from equilibrium effects. The most popular
such novel feature is the phenomenon of {\it Stochastic Resonance}\/
\cite{rmp98}: It refers to the fact that
an optimal level of applied or intrinsic noise
can dramatically boost the response (or, more generally
the transport) to typically weak, time-dependent input signals
in nonlinear stochastic systems.
This theme naturally plays a crucial role in biology with its variety of
threshold-like systems that are subjected to
noise influences \cite{srbiology}.

A more recent but increasingly popular example of the constructive role of
fluctuations (intrinsic and external, alike)
is the noise-assisted transport in periodic systems,
namely the so-called {\it Brownian Motors}\/  \cite{brownianmotors}.

Both topics are still very much active: This Focus issue contains both, experimental
and theoretical contributions to Stochastic Resonance by
Bechinger {\it et al} \/ \cite{bechinger05},
Casado-Pascual {\it et al} \/ \cite{casado05}, and Gammaitoni {\it et al}
\cite{bulsara05}. The theme of noise-assisted transport
is  multifaceted and very rich; this is corroborated with  several appealing
contributions by  Linke {\it et al} \/ \cite{linke05},
Borromeo and Marchesoni\/
\cite {borromeo05}, Savel'ev and Nori \cite{savelev05}
and Eichhorn {\it et al} \/ \cite{eichhorn05}.

\section{resume}
This Focus issue on ``100 Years of Brownian Motion'' is not only
timely but also circumstantiates that research in this area is very
much alive and still harbors  plenty of surprises that only wait to
get unravelled by future researchers. The original ideas that
Einstein put forward in 1905 are very modern and still find their
way to applications in such diverse areas as soft matter physics
\cite{frey05}, including the granular systems investigated here by
Brilliantov and P\"oschel \cite{poeschel05} and the soliton
diffusion in linear defects \cite {borromeo05,solitons}, solid state
physics, chemical physics, computational physics, and beyond. In
recent years, ideas and tools developed within the context of the
Brownian motion theory are gaining increasing impact in life
sciences (the contribution by Zaks {\it et al.} \cite{zaks05}
provides a timely example) and even in human studies, where
econophysics is becoming a lively crossroad of interdisciplinary
research, as shown with the study by Bouchaud  \cite{bouchaud05} in
this issue.

We Guest Editors  share the
confident belief that the contributions in this Focus issue by leading
practitioners from a
diverse range of backgrounds will together provide a fair and accurate
snapshot of the current state of this rich and interdisciplinary research field.
Last but not least, we hope that this collection of articles will
stimulate readers into pursuing future research of their own.

\begin{acknowledgments}
PH gratefully acknowledges financial support by the DAAD-KBN
(German-Polish project \textit{Stochastic Complexity}),  the Deutsche
Forschungsgemeinschaft via grant HA 1517/13-4 and the collaborative research
grants SFB~486 and SFB~631. PH also likes to thank Hajo Leschke for bringing the
work of Sutherland, 1859--1911, on the Einstein relation to his attention.
\end{acknowledgments}

\end{document}